\documentclass[12pt]{article}
\usepackage{graphicx,color,latexsym,multirow,url,natbib}
\usepackage{ifthen,pifont,comment,graphicx,amssymb,amsthm,amsmath,enumerate}
\usepackage[dvipsnames]{xcolor}
\usepackage[letterpaper, left=3cm, top=3cm, right=3cm,
bottom=3cm]{geometry}
\usepackage{charter}
\RequirePackage[colorlinks,citecolor=blue]{hyperref}

 \makeatletter
\renewcommand\section{\@startsection {section}{1}{\z@}
                                   {-3.5ex \@plus -1ex \@minus -.2ex}
                                    {2.3ex \@plus.2ex}
                                   {\centering\normalsize\scshape}}
\renewcommand\subsection{\@startsection {subsection}{1}{\z@}
                                   {-3.5ex \@plus -1ex \@minus -.2ex}
                                   {2.3ex \@plus.2ex}
                                   {\centering\normalsize}}

\def\diag{{\rm diag}}
\def\log{{\rm log}}

\def\tr{{\rm tr}}
\def\ud{{\rm d }}

\def\bD {\mathbf{D}}

\def\bS {\mathbf{S}}
\def\bs {\mathbf{s}}

\def\bu {\mathbf{u}}

\def\bV {\mathbf{V}}

\def\bY {\mathbf{Y}}

\newcommand{\bfb}   {\mbox{\boldmath $\beta$}}

\newcommand{\bftau} {\mbox{\boldmath $\tau$}}

\newcommand{\bfSigma} {\mbox{\boldmath $\Sigma$}}
\newcommand{\bfsigma} {\mbox{\boldmath $\sigma$}}
\makeatother

\begin{document}

\begin{center}
{\large\bf Two New Algorithms for Solving Covariance Graphical Lasso Based on Coordinate Descent and ECM} \\
\vspace{.3in}
{ Hao Wang} \\
\medskip
{\it Department of Statistics, University of South Carolina, \\ Columbia, South Carolina ~29208, U.S.A.} \\
 haowang@sc.edu

\bigskip
(First version: May 17, 2012)
\\
\end{center}
\begin{abstract}
Covariance graphical lasso applies a lasso penalty on the elements of the covariance matrix.  This method is useful because it not only produces sparse estimation of covariance matrix but also discovers marginal independence structures by generating zeros in the covariance matrix.  We propose and explore two new algorithms for solving the covariance graphical lasso problem. Our new algorithms are based on coordinate descent and ECM.  We show that these two algorithms are more attractive than the only existing competing algorithm of \citet{BienTib2011} in terms of simplicity, speed and stability. We also discuss convergence properties of our algorithms.
\end{abstract}

\bigskip

\section{Introduction}
\citet{BienTib2011} proposed a covariance graphical lasso procedure for simultaneously estimating covariance matrix and marginal dependence structures.  Let $\bS$ be the sample covariance matrix
such that  $\bS=\bY^\prime \bY/n$ where $\bY (n\times p)$ is the data
matrix of $p$ variables and $n$ samples.  A basic version of their covariance graphical lasso problem is to minimize the following objective function:
\begin{eqnarray} \label{eq:obj}
g(\bfSigma) = \log(\det \bfSigma)+\tr(\bS\bfSigma^{-1}) + \rho
||\bfSigma||_1,
\end{eqnarray}
over the space of positive definite matrices $M^+$ with $\rho\geq 0$
being the shrinkage parameter. Here, $\bfSigma=(\sigma_{ij})$ is the
$p\times p$ covariance matrix and  $||\bfSigma||_1 =
\sum_{1\leq i,j \leq p} |\sigma_{ij}|$ is the $L_1$-norm of
$\bfSigma$. A general version of the covariance graphical lasso in \citet{BienTib2011} allows different shrinkage parameters for different elements in $\bfSigma$. To ease exposition, we describe our methods in the context of the simple case of common shrinkage parameter in (\ref{eq:obj}). All of our results can be extended to the general version of different shrinkage parameters with little difficulty.

Because of the $L_1$-norm term, the covariance graphical lasso is possible to set some of the off-diagonal elements of $\bfSigma$ exactly equal to zero in its minimum point of (\ref{eq:obj}). Zeros in $\bfSigma$ encode marginal independence structures among the components of a multivariate normal random vector with covariance matrix $\bfSigma$. It is distinctly different from the concentration graphical models (also referred to as covariance selection models due to \citealt{Dempster72}) where zeros are in the concentration matrix $\bfSigma^{-1}$ and are associated with conditional independence.

The objective function (\ref{eq:obj}) is not convex, imposing computational challenges for minimizing it.  \citet{BienTib2011} proposed a complicated majorize-minimize approach that iteratively solves convex approximation to the original nonconvex problem.  In the current paper, we develop two alternative algorithms for minimizing (\ref{eq:obj}): the coordinate descent algorithm and the Expectation/Conditional Maximization (ECM) algorithm.  We discuss their convergence properties and investigate their computational efficiency through simulation studies.  In comparison with \citet{BienTib2011}'s algorithm, our new algorithms are much simpler to implement, substantially faster to run and  numerically more stable in our tests.

Neither the coordinate descent algorithm nor the ECM algorithm is new to the model fitting for regularized problems. The coordinate descent algorithm is shown to be very competitive for fitting convex and even some non-convex penalized regression models \citep{FHHT2007,WuLange2008,BrehenyHuang2011} as well as the  concentration graphical models \citep{Friedman08}.  The ECM algorithm (or its variants) has been developed in the Bayesian framework for finding the maximum a posteriori (MAP) estimation
of regularized linear regression models \citep{PolsonScott2011,Armagan11}.  However, our application of these two algorithms to covariance graphical lasso models is new  and unexplored before. In this sense, our work documents that the coordinate descent algorithm and the ECM algorithm are also quite powerful in solving covariance graphical lasso models.

\section{Two proposed algorithms}
\subsection{Coordinate descent algorithm} \label{sec:cd}
Our first algorithm to minimize the objective function (\ref{eq:obj}) uses the simple idea of coordinate descent methods.   We show how to update $\bfSigma$ one column and row at a time while holding all of the rest elements in $\bfSigma$ fixed. Without loss of generality, we focus on the last column and row. Partition $\bfSigma$ and $\bS$ as follows:
\begin{equation}\label{eq:partition}
\bfSigma = \left(
\begin{array}{cc}
\bfSigma_{11},\bfsigma_{12} \\ \bfsigma_{12}^\prime,\sigma_{22}
\end{array} \right ),\quad \bS = \left(
\begin{array}{cc}
\bS_{11},\bs_{12} \\ \bs_{12}^\prime,s_{22}
\end{array} \right ),
 \end{equation}
 where \textit{(a)} $\bfSigma_{11}$ and $\bS_{11}$ are the covariance matrix  and the sample covariance matrix of the first $p-1$ variables, respectively; \textit{(b)}  $\bfsigma_{12}$ and $\bs_{11}$ are the covariances and the sample covariances between the first $p-1$ variables and the last variable, respectively; and \textit{(c)} $\bfsigma_{22}$ and $\bs_{ss}$ are the variance  and the sample variance of the last variable, respectively.

 Let $$\bfb=\bfsigma_{12}, \quad \gamma = \sigma_{22} - \bfsigma_{12}^\prime \bfSigma_{11}^{-1} \bfsigma_{12},$$  and apply the block matrix inversion to $\bfSigma$ using blocks $(\bfSigma_{11},\bfb,\gamma)$:
\begin{equation}\label{eq:blockinv}
\bfSigma^{-1} = \left(
\begin{array}{cc}
\bfSigma_{11}^{-1}+\bfSigma_{11}^{-1} \bfb \bfb^\prime\bfSigma_{11}^{-1}\gamma^{-1} & -\bfSigma_{11}^{-1} \bfb\gamma^{-1} \\ -\bfb^\prime \bfSigma_{11}^{-1}\gamma^{-1} & \gamma^{-1}
\end{array} \right ).
 \end{equation}
After removing some constants not involving $(\bfb,\gamma)$, the three terms in (\ref{eq:obj}) can be expressed as a function of $(\bfb,\gamma)$:
\begin{eqnarray}
\log(\det\bfSigma) &= & \log(\gamma), \nonumber \\
\tr(\bS\bfSigma^{-1}) & = & \bfb^\prime \bfSigma_{11}^{-1} \bS_{11} \bfSigma_{11}^{-1} \bfb\gamma^{-1}-2\bs_{12}^\prime\bfSigma_{11}^{-1}\bfb\gamma^{-1}+s_{22}\gamma^{-1}, \nonumber
\\ \rho ||\bfSigma||_1& = & 2 \rho ||\bfb||_1 + \rho (\bfb^\prime \bfSigma_{11}^{-1} \bfb +\gamma),
\end{eqnarray}
which lead to the following objective function with respect to $(\bfb,\gamma)$:
\begin{equation}\label{eq:betagamma}
\min_{\bfb,\gamma}\biggr\{ \log(\gamma) + \bfb^\prime \bfSigma_{11}^{-1} \bS_{11} \bfSigma_{11}^{-1} \bfb\gamma^{-1}-2\bs_{12}^\prime \bfSigma_{11}^{-1}\bfb\gamma^{-1} + s_{22}\gamma^{-1} + 2 \rho |\bfb| + \rho \bfb^\prime \bfSigma_{11}^{-1} \bfb + \rho \gamma \biggr\}.
\end{equation}

For $\gamma$, removing terms in (\ref{eq:betagamma}) that do not depend on $\gamma$  gives
$$\min_{\gamma} \biggr\{ \log(\gamma) + a \gamma^{-1} +  \rho \gamma \biggr\},$$
where $a=\bfb^\prime \bfSigma_{11}^{-1} \bS_{11} \bfSigma_{11}^{-1} \bfb - 2\bs_{12}\bfSigma_{11}^{-1}\bfb +s_{22}$. Clearly, it is solved as follows:
\begin{equation}\label{eq:gammacd}
\hat{\gamma} = \left\{
\begin{array}{cl}
a & \text{if } \rho=0,\\
(-1+\sqrt{1+4a\rho})/ (2\rho) & \text{if } \rho \neq 0.
\end{array} \right.
\end{equation}

For $\bfb$, removing terms in (\ref{eq:betagamma}) that do not depend on $\bfb$  gives
\begin{align}\label{eq:betaobj}
\min_{\bfb} \biggr\{ \bfb^\prime \bV \bfb-2 \bu^\prime \bfb +2\rho||\bfb||_1 \biggr\},\end{align}
where $\bV =(v_{ij})= \bfSigma_{11}^{-1} \bS_{11} \bfSigma_{11}^{-1}\gamma^{-1}+\rho\bfSigma_{11}^{-1}$, $\bu=\bs_{12}^\prime \bfSigma_{11}^{-1}\gamma^{-1}.$  The problem in (\ref{eq:betaobj}) is a lasso problem and can be efficiently solved by fast coordinate descent algorithms \citep{FHHT2007,WuLange2008}.  Specifically, for $j \in
\{1,\ldots, p-1\}$, the minimum point of  $(\ref{eq:betaobj})$ along the coordinate direction in which $\beta_j$ varies is:
\begin{align}\label{eq:betasoftthr}
\hat{\beta}_j & = \mathcal{S}(u_j - \sum_{k\neq j} v_{kj} \hat{\beta}_k,\rho)/v_{jj},
\end{align}
where $\mathcal{S}$ is the soft-threshold operator: $$\mathcal{S}(x,t) = \textrm{sign}(x)(|x|-t)_+.$$
The update (\ref{eq:betasoftthr}) is iterated for $j=1,\ldots,p-1,1,2,\ldots,$ until convergence. We then update the column as $(\bfsigma_{12}=\bfb,\sigma_{22}=\gamma + \beta^\prime \bfSigma_{11}^{-1} \beta)$ followed by cycling through all columns until convergence.  This algorithm can been viewed as a block coordinate descent method with $p$ blocks of $\bfb$s and another $p$ blocks of  $\gamma$s. The algorithm is summarized as follows: \\ \\
\textbf{Coordinate descent algorithm  } Given input $(\bS,\rho)$, start with $\bfSigma^{(0)}$, at the $(k+1)$th iteration ($k=0,1,\ldots$)
\begin{enumerate}[1.]
\item Let $\bfSigma^{(k+1)}=\bfSigma^{(k)}$.
\item
For $i=1,\ldots,p$,
\begin{enumerate}[(a)]
\item Partition $\bfSigma^{(k+1)}$ and $\bS$ as in (\ref{eq:partition}).
\item Compute $\gamma$ as in (\ref{eq:gammacd}).
\item Solve the lasso problem (\ref{eq:betaobj}) by repeating $(\ref{eq:betasoftthr})$ until convergence.
\item Update $\bfsigma_{12}^{(k+1)}=\beta,\bfsigma_{21}^{(k+1)} =\beta^\prime$, $\sigma_{22}^{(k+1)} = \gamma + \beta^\prime \bfSigma_{11}^{-1} \beta$.
\end{enumerate}
\item Let $k=k+1$ and repeat (1)-(3) until convergence.
\end{enumerate}

\subsection{ECM algorithm}\label{sec:ecm}
The second algorithm to find the estimator $\hat{\bfSigma} = \textrm{argmin}_{\bfSigma \in M^+}g(\bfSigma) $  utilizes representation of the solution to (\ref{eq:obj}) as a MAP estimator $\hat{\bfSigma}$ for the posterior distribution:
$p(\bfSigma \mid \bS) \propto \exp\{ -{1\over 2}g(\bfSigma)\}.$  We propose to use the ECM algorithm for iterative
compute its modes.  The key to the ECM algorithm is the choice of latent variables.  Note that, for any off-diagonal element $\sigma_{ij}$, we have the following well-known representation:
\begin{equation}\label{eq:scalerep}
{\rho\over 2}\exp\{ - \rho |\bfsigma_{ij}| \} =  \int_{0}^\infty  (2\pi\tau_{ij})^{-{1\over 2}}\exp(-{\sigma^2_{ij}\over 2\tau_{ij}}) {\rho^2\over2}\exp(-{\rho^2\over 2} \tau_{ij} ) \ud \tau_{ij},
\end{equation}
where $\tau_{ij}$ the latent scale parameter. This suggests that the posterior distribution $p(\bfSigma \mid \bS) \propto \exp\{ -{1\over 2}g(\bfSigma)\}$ is the marginal distribution of the following complete posterior distribution of $(\bfSigma,\bftau)$ where $\bftau=(\tau_{ij})_{i<j}$:
\begin{align}\label{eq:jointpost}
p(\bfSigma ,\bftau \mid \bS) \propto |\bfSigma|^{-{1\over 2}}\exp\{-{1\over 2} \tr(\bS\bfSigma) \}\prod_{i<j} \bigg \{ \tau_{ij}^{-{1\over 2}}\exp(-{\sigma^2_{ij}\over 2\tau_{ij}}) \exp(-{\rho^2\over 2} \tau_{ij} ) \bigg \} \nonumber \\ \times \prod_{i=1}^p \bigg\{\exp(-{{\rho \over 2}} \sigma_{ii}) \bigg\}.
\end{align}
Consequently, the covariance graphical lasso estimator from (\ref{eq:obj}) is equivalent to the marginal mode for $\bfSigma$ in (\ref{eq:jointpost}). We now describe how to compute this marginal mode for $\bfSigma$ using the ECM algorithm \citep{MengRubin1993}.

To perform the E-step, we calculate the expected complete log-posterior by replacing $\tau_{ij}^{-1}$ with their conditional expectations $\textsc{E}(\tau_{ij}^{-1} \mid \bS,\bfSigma^{(k)})$ given the current $\bfSigma^{(k)}$. Following the standard results of the inverse-Gaussian distribution, we have
$$\textsc{E}({1\over \tau_{ij}} \mid \bS,\bfSigma^{(k)}) = {\rho \over |\sigma^{(k)}_{ij}|},$$ which leads to the following criterion function after removing the terms that do not depend on $\bfSigma$,
\begin{align}\label{eq:Estep}
Q(\bfSigma \mid \bfSigma^{(k)}) &= \int \log p(\bfSigma, \bftau \mid \bS) p(\bftau \mid \bfSigma^{(k)},\bS)\ud \bftau \nonumber \\
&\propto -\log(\det \bfSigma)-\tr(\bS\bfSigma^{-1})-{\rho} \sum_{i<j} {\sigma_{ij}^2\over |\sigma_{ij}^{(k)}|} -\rho \sum_{i} \sigma_{ii}.
\end{align}

The CM-step then maximizes (\ref{eq:Estep}) along each column (row) of $\bfSigma$. Again, without loss of generality, we focus on the last column and row. Partition $\bfSigma$ and $\bS$ as in (\ref{eq:partition}) and consider the same transformation from $(\bfsigma_{12},\sigma_{22})$ to $(\bfb = \bfsigma_{12},\gamma =\sigma_{22} - \bfsigma_{12}^\prime \bfSigma_{11}^{-1} \bfsigma_{12})$. After dropping of terms not involving $(\bfb,\gamma)$, the four terms in (\ref{eq:Estep}) can be written as functions of $(\bfb,\gamma)$
\begin{align*}
\log(\det\bfSigma) &= \log(\gamma)\\
\tr(\bS\bfSigma^{-1}) & = \bfb^\prime \bfSigma_{11}^{-1} \bS_{11} \bfSigma_{11}^{-1}. \bfb\gamma^{-1}-2\bs_{12}^\prime\bfSigma_{11}^{-1}\bfb\gamma^{-1}+s_{22}\gamma^{-1},
\\\rho \sum_{i<j} {\sigma_{ij}^2\over |\sigma_{ij}^{(k)}|} & = {\rho}  \bfb^\prime \bD^{-1} \bfb,
\\  \rho \sum_{i} \sigma_{ii} & = \rho (\bfb^\prime \bfSigma_{11}^{-1} \bfb +\gamma).
\end{align*}
where  $\bD = \diag(|\bfsigma^{(k)}_{12}|)$. Holding all but $(\bfb,\gamma)$ fixed, we can then rewrite (\ref{eq:Estep}) as
\begin{align}\label{eq:CMstep}
Q(\bfb,\gamma \mid \bfSigma^{(k)}) =  -\biggr \{ \log(\gamma) + \bfb^\prime \bfSigma_{11}^{-1} \bS_{11} \bfSigma_{11}^{-1} \bfb\gamma^{-1}-2\bs_{12}^\prime \bfSigma_{11}^{-1}\bfb\gamma^{-1} + s_{22}\gamma^{-1} \nonumber \\ + \rho \bfb^\prime \bD^{-1} \bfb + \rho \bfb^\prime \bfSigma_{11}^{-1} \bfb + \rho \gamma \biggr \}.
\end{align}
For $\gamma$, it is easy to derive from (\ref{eq:CMstep}) that  the conditional maximum point given $\bfb$ is the same as in (\ref{eq:gammacd}). On the other hand, given $\gamma$, (\ref{eq:CMstep}) can be written as a function of $\bfb$,
$$ Q(\bfb \mid \gamma, \Sigma^{(k)}) = -\biggr\{ \bfb^\prime (\bV+\rho\bD^{-1}) \bfb-2 \bu^\prime \bfb \biggr\},$$
where $\bV$ and $\bu$ are defined in (\ref{eq:betaobj}). This implies that the conditional maximum point of $\bfb$ is \begin{equation}\label{eq:betaecm} \bfb = (\bV+\rho \bD^{-1})^{-1} \bu. \end{equation}
Cycling through every column generates a sequence of CM steps.  The total number of CM-steps is $2p$ because there are $p$ columns and for each column the CM involves two steps: one for $\bfb$ and the other for $\gamma$. The ECM algorithm can be summarized as follows: \\ \\
\textbf{ECM algorithm} Given input $(\bS,\rho)$, start with $\bfSigma^{(0)}$, at the $(k+1)$th iteration, $k=0,1,\ldots,$
\begin{enumerate}[1.]
\item Let $\bfSigma^{(k+1)}=\bfSigma^{(k)}$.
\item \textbf{E-step: } Compute $Q$ as in (\ref{eq:Estep}).
\item \textbf{CM-step: }
For $i=1,\ldots,p$,
\begin{enumerate}[(a)]
\item Partition $\bfSigma^{(k+1)}$ and $\bS$ as in (\ref{eq:partition}).
\item Compute $\gamma$ as in (\ref{eq:gammacd}) and $\bfb$ as in (\ref{eq:betaecm}).
\item Update $\bfsigma_{21}^{(k+1)}=\beta,\bfsigma_{12}^{(k+1)} =\beta^\prime$, $\sigma_{22}^{(k+1)} = \gamma + \beta^\prime \bfSigma_{11}^{-1} \beta$.
\end{enumerate}
\item Let $k=k+1$ and repeat (1)-(4) until convergence.
\end{enumerate}

\subsection{Algorithm convergence}
The convergence of the proposed (block) coordinate descent algorithm can be addressed by the theoretical innovation for block coordinate descent methods for non-differentiable minimization by \citet{Tseng2001}. The key to the application of the general theory there to our algorithm is the separability of the non-differentiable penalty terms in (\ref{eq:obj}).  First, from (\ref{eq:gammacd}) and (\ref{eq:betasoftthr}), the objective function $g$ has a unique minimum point in each coordinate block.  This satisfies the conditions of Part (c) of Theorem 4.1 in \citet{Tseng2001} and hence implies that the algorithm converges to a coordinatewise minimum point. Second, because all directional derivatives exist,  by Lemma 3.1 of \citet{Tseng2001}, each coordinatewise minimum point is a stationary point.   A close-related argument has been given by \citet{BrehenyHuang2011} to show the convergence of coordinate decent algorithm for nonconvex penalized regression models.

The convergence of the proposed ECM algorithm follows directly from the results of \citet{MengRubin1993}. First, the analytical solutions (\ref{eq:gammacd}) and (\ref{eq:betaecm}) of CM-steps are unique for any $|\sigma^{(k)}_{ij}| \neq 0 $.  Second, it can be easily seen that the construction of  the $2p$ CM-steps satisfies the ``space filling'' condition for any $\bfSigma$ because the CM-steps cycle through the whole parameter space.  Thus, the two conditions of Theorem 3 of \citet{MengRubin1993} hold for the proposed ECM and so all limiting points of the ECM sequence $\{\bfSigma^{(k)}\}$ are stationary points of (\ref{eq:obj}).

Using the latent variables $\bftau$ makes the EM-type optimization simpler than direct optimization, however, it has some issues.  The ECM must be initialized at $|\sigma^{(0)}_{ij}| \neq 0$ for all $i<j$. Otherwise, from (\ref{eq:Estep}), the criterion function $Q(\bfSigma \mid \bfSigma^{(k)}) =\infty$, the algorithm will then stuck. Another related issue is that, although the sequence $\{\sigma^{(k)}_{ij}\}$ will possibly converge to zero, they cannot be identically equal to zero.   We want to point out that these issues are not unique to the covariance graphical models. Existing EM-type algorithms based on the latent scale parameters for fitting regularized linear models (e.g., \citealt{PolsonScott2011,Armagan11}) have the same problems.

\section{Comparison of algorithms}
We compare the computational aspects of the two
algorithms in Section \ref{sec:cd} and \ref{sec:ecm} with \citet{BienTib2011}'s algorithm.  We consider two scenarios:
\begin{itemize}
\item  A \textit{sparse} model taken from  \citet{BienTib2011} with $\sigma_{i,i+1}=\sigma_{i,i-1}=0.4, \sigma_{ii}=\delta$  and zero otherwise. Here, $\delta$ is chosen so that the condition number of $\Sigma$ is $p$.
\item A \textit{dense} model with $\sigma_{ii}=2$ and $\sigma_{ij}=1$ for $i\neq j$.
\end{itemize}

The algorithm of \citet{BienTib2011} is coded in R with its built-in functions.  To be comparable to it, we implemented the ECM and the coordinate descent algorithms in R without writing any functions in a compiled language.   All
computations were done on a Intel Xeon X5680 3.33GHz processor.

For either the sparse or the dense model, we first generated two datasets of dimension
$(p,n)=(100,200)$ and $(p,n)=(200,400)$, respectively. Then, for each of the four datasets,
we applied the three algorithms for a range of $\rho$ values. All computations were initialized at the sample covariance matrix, i.e., $\bfSigma^{(0)}=\bS$. For \citet{BienTib2011}'s algorithm, we followed the default setting of tuning parameters provided by the ``spcov'' package (\url{http://cran.r-project.org/web/packages/spcov/index.html}). For the ECM and the coordinate descent algorithms, we used the same criterion as \citet{BienTib2011}'s algorithm to stop the iterations:  The procedure stops when the change of the objective function is less than $10^{-3}$.

First, we  report the performance of numerical stability.  In the case of sparse models and $p=200$, \citet{BienTib2011}'s algorithm would not run. A careful inspection of the algorithm details  reveals that the Newton-Raphson step there to find $\delta_-$ fails in this case.  The same problem occurs quite often when we apply \citet{BienTib2011}'s algorithm to different datasets generated from different $\bfSigma$.  This suggests that \citet{BienTib2011}'s algorithm may lack numerical stability. It may be possible to reduce this numerical problem by calibrating some of \citet{BienTib2011}'s tuning parameters such as the initial value for the Newton-Raphson method.  In any event, it may be safe to conclude that  \citet{BienTib2011}'s algorithm requires either very careful tuning or lacks stability in certain cases.

\begin{figure}[htbp]
\center
\scriptsize
\begin{tabular}{cc}
   (a) & (b) \\
   \includegraphics[width=2.7in]{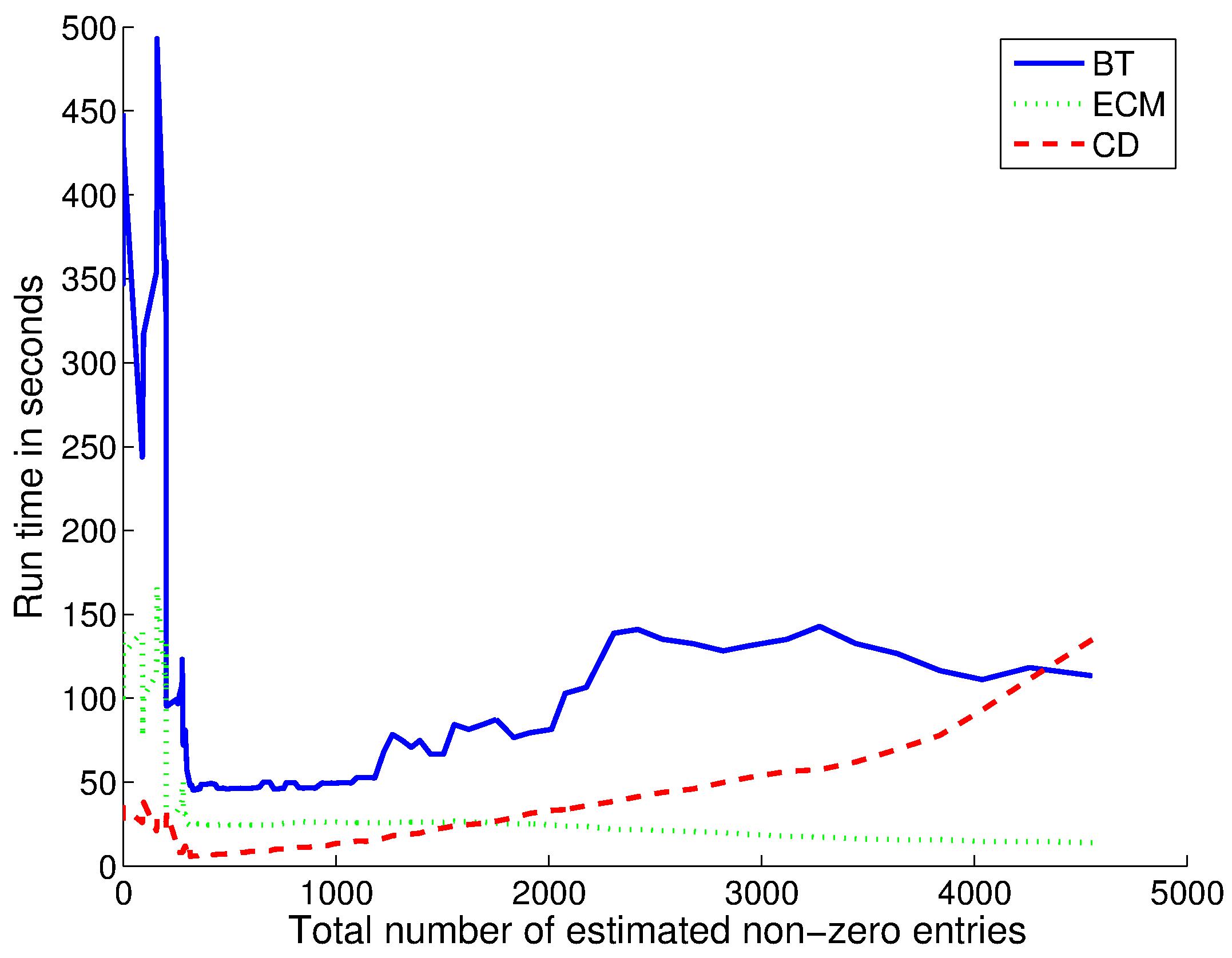}&
   \includegraphics[width=2.7in]{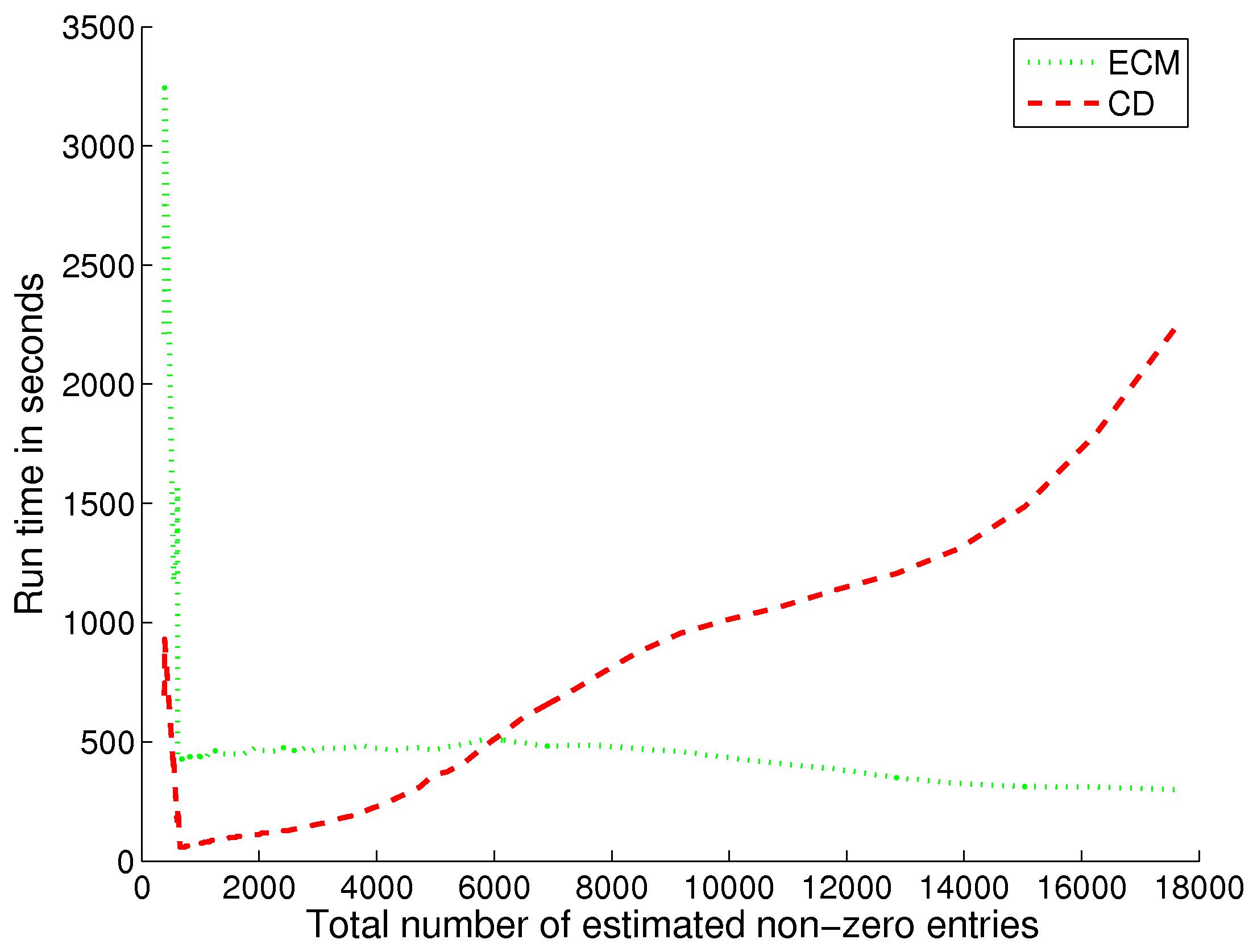}\\
   & \\
   & \\
   & \\
      (c) & (d) \\
    \includegraphics[width=2.7in]{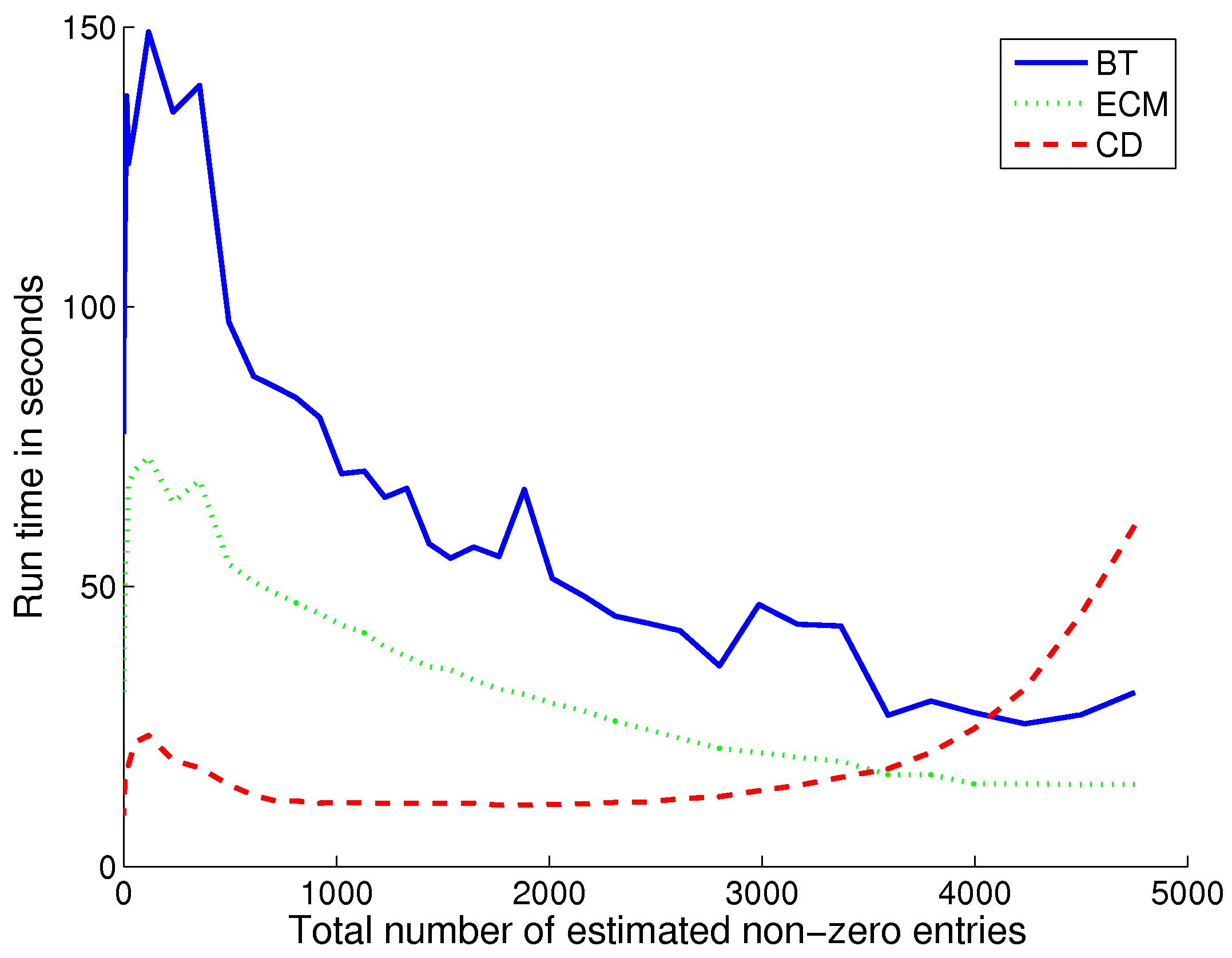}&
   \includegraphics[width=2.7in]{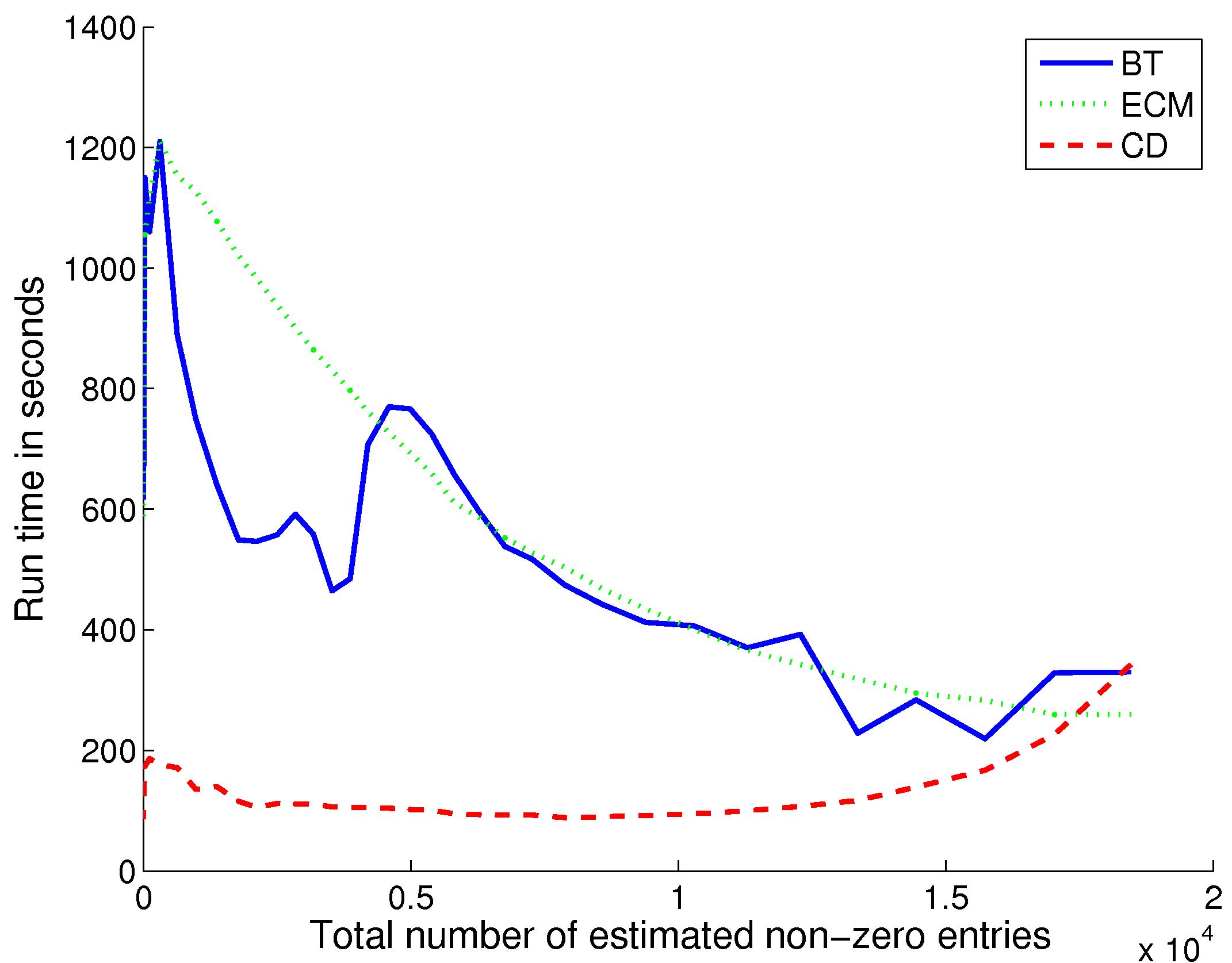}
 \end{tabular}
 \caption{CPU time (in seconds) is plotted against the number of non-zero edges estimated for four different models and three different algorithms. The four models are: sparse $\bfSigma$ and $p=100$ (Panel a); sparse $\bfSigma$ and $p=200$ (Panel b); dense $\bfSigma$ and $p=100$ (Panel c), and dense $\bfSigma$ and $p=200$ (Panel d). The three algorithms are: \citet{BienTib2011} (BT, solid line), ECM (dotted line) and coordinate descent (CD, dashed line).  Each computation is initialized at the sample covariance matrix $\bfSigma^{(0)}=\bS$.
 } \label{fig:time}
\end{figure}

Now, we compare the computing speed.  The four panels in Figure \ref{fig:time} display the CPU times of the three algorithms for each of the four datasets respectively. CPU time in seconds is plotted against the total number of off-diagonal non-zero elements estimated by the coordinate descent algorithm. The ECM and the coordinate descent algorithms are much faster than the \citet{BienTib2011}'s algorithm except for the dense model and $p=200$ in which the ECM is comparable with \citet{BienTib2011}. Moreover, the coordinate descent algorithm seems to be particularly attractive for larger $\rho$ as its run time generally decreases when the total number of estimated non-zero elements decreases. 

\begin{figure}[htbp]
\center
\scriptsize
\begin{tabular}{cc}
   (a) & (b) \\
   \includegraphics[width=2.7in]{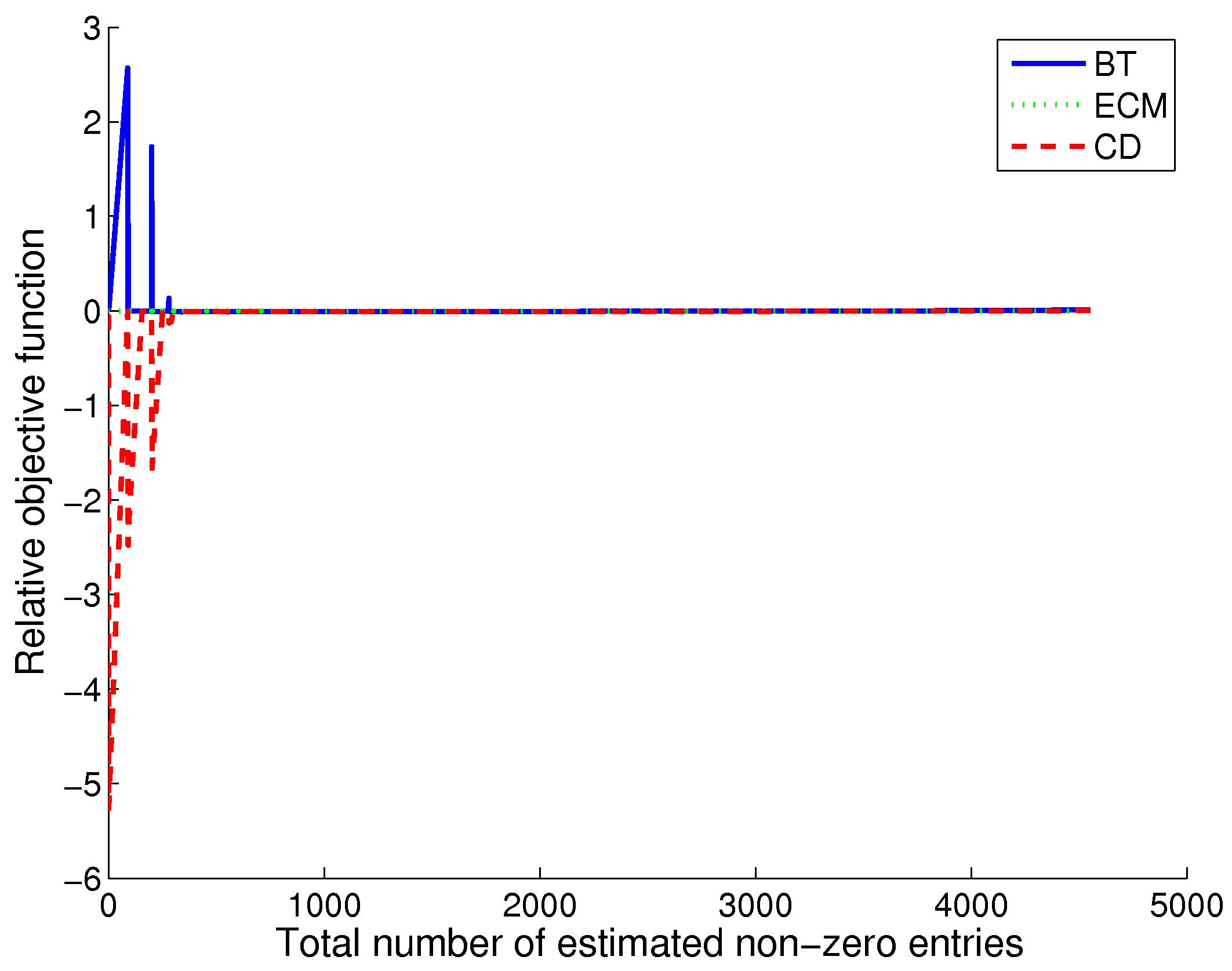}&
   \includegraphics[width=2.7in]{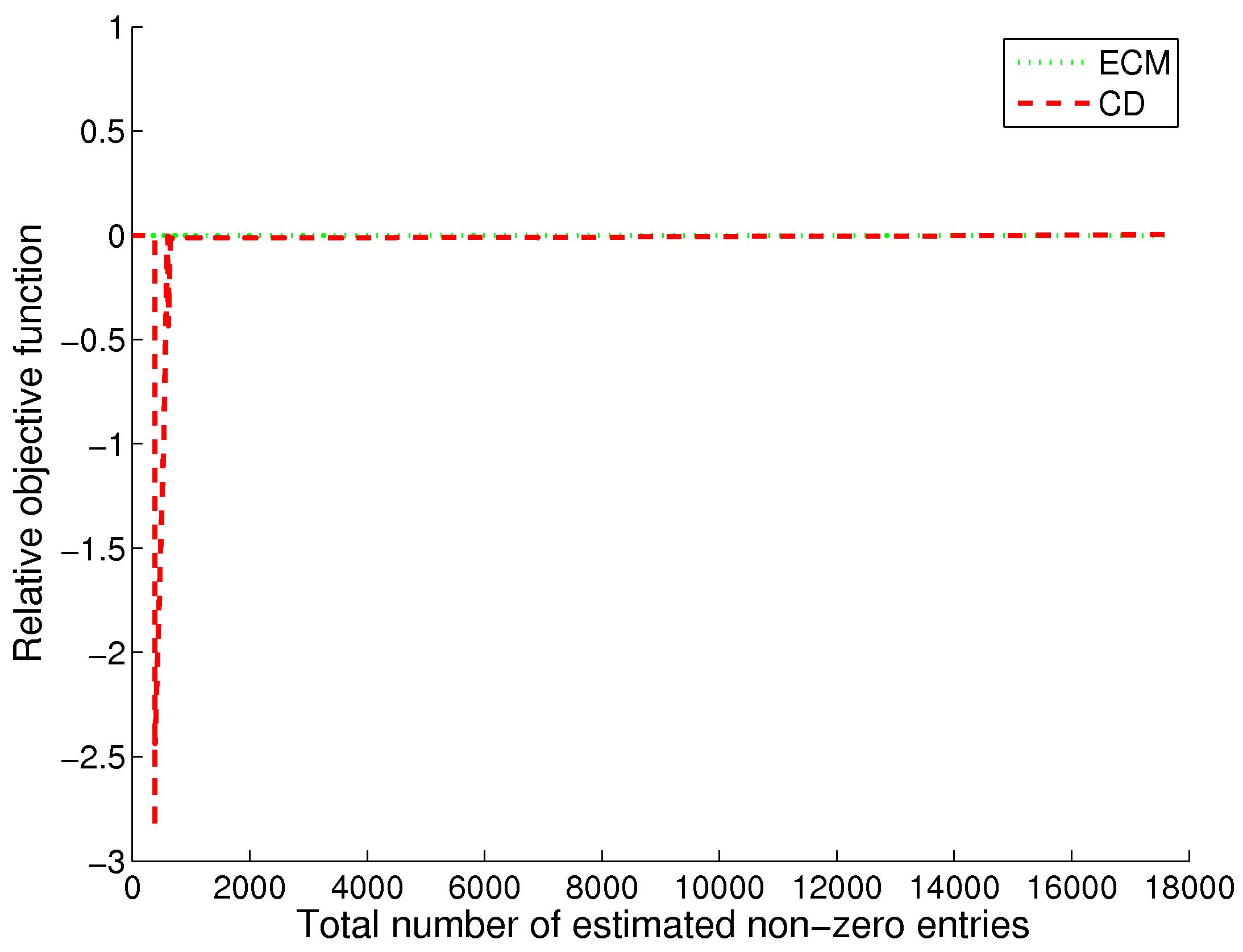}\\
   & \\
   & \\
   & \\
      (c) & (d) \\
    \includegraphics[width=2.7in]{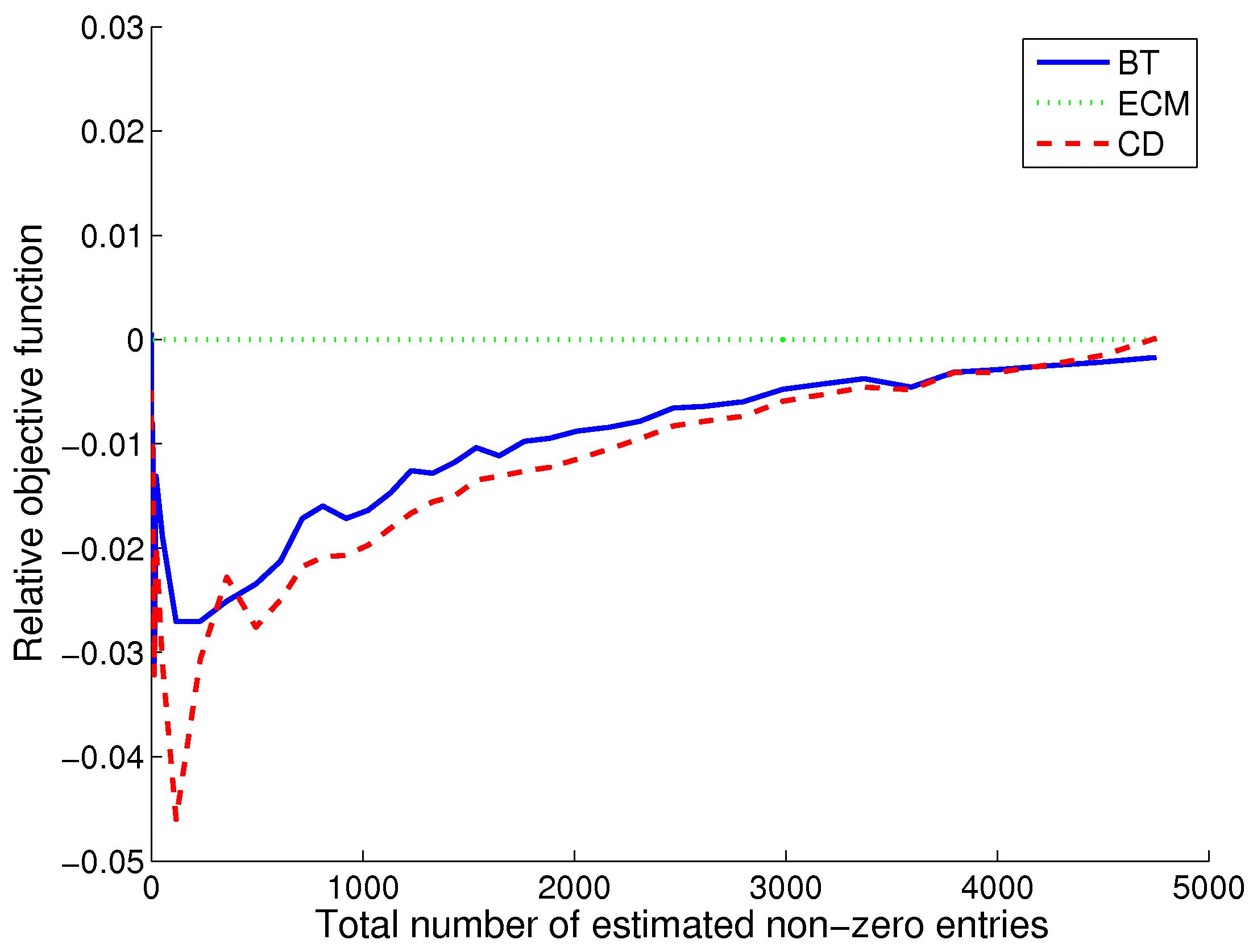}&
   \includegraphics[width=2.7in]{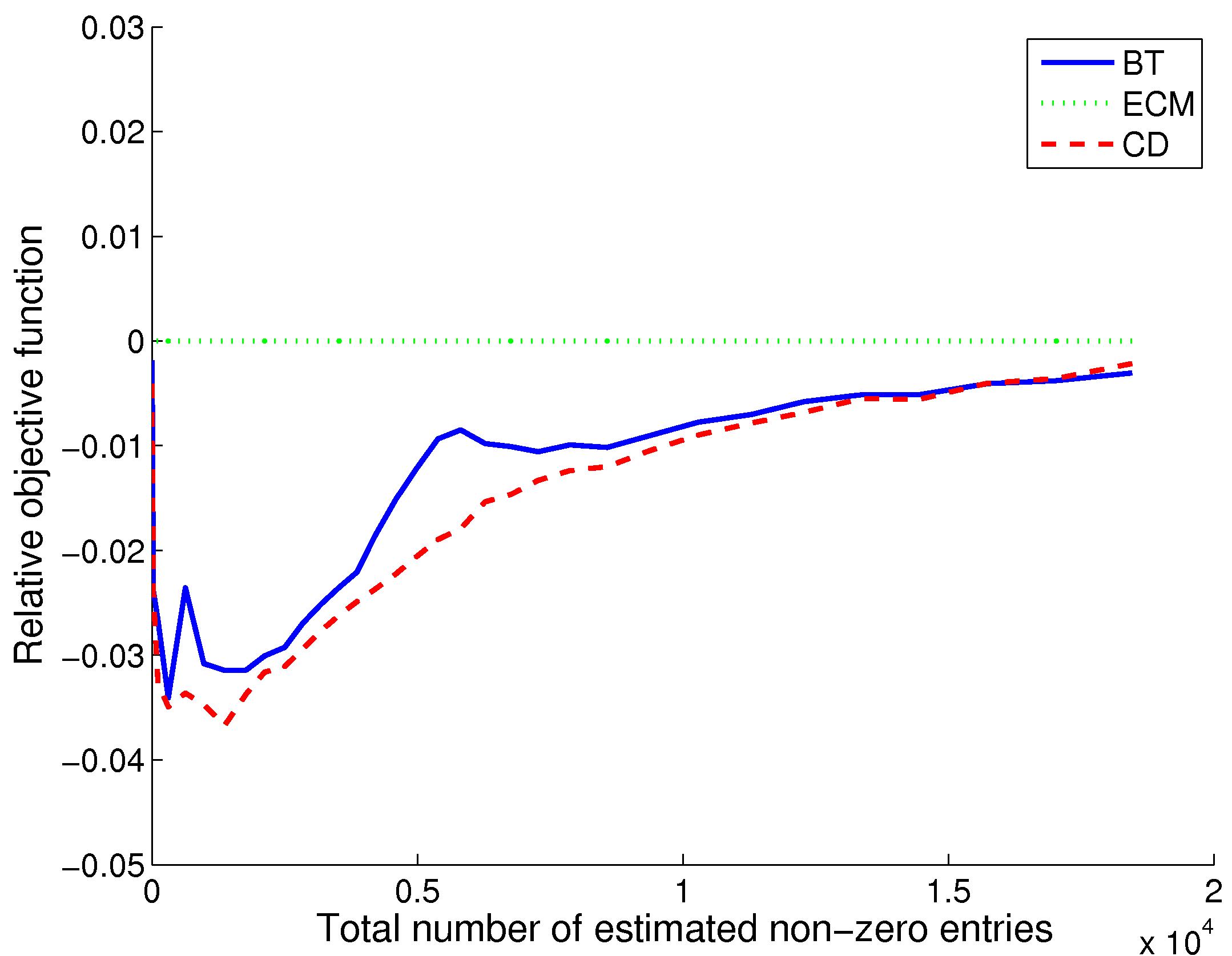}
 \end{tabular}
 \caption{Relative minimum value of objective function, defined in (\ref{eq:relobj}), is plotted against the number of estimated non-zero elements for four different models and three different algorithms. The four models are: sparse $\bfSigma$ and $p=100$ (Panel a); sparse $\bfSigma$ and $p=200$ (Panel b); dense $\bfSigma$ and $p=100$ (Panel c), and dense $\bfSigma$ and $p=200$ (Panel d). The three algorithms are: \citet{BienTib2011} (BT, solid line), ECM (dotted line) and coordinate descent (CD, dashed line).  Each computation is initialized at the sample covariance matrix $\bfSigma^{(0)}=\bS$.} \label{fig:obj}
\end{figure}

Next, we examine the ability of the algorithms to search for minimum points. To do so, we compared the minimum values of the objective functions achieved by each algorithm.  For each dataset and each $\rho$,  We calculated relative minimum values of objective functions defined as:
\begin{align}\label{eq:relobj}
g(\hat{\bfSigma}_{BT})-g(\hat{\bfSigma}_{ECM}),\quad g(\hat{\bfSigma}_{CD})-g(\hat{\bfSigma}_{ECM}),
\end{align}
where $\hat{\bfSigma}_{BT}$, $\hat{\bfSigma}_{CD}$ and $\hat{\bfSigma}_{ECM}$ are the minimum points found by \citet{BienTib2011}, the coordinate descent and the ECM algorithms, respectively. Thus, a negative value
indicates that the algorithm finds better points than the ECM procedure and
a smaller relative objective function indicates a better relative performance of the method.  The four panels in Figure \ref{fig:obj} display the relative minimum values of the objective functions for each of the four datasets, respectively.  The relative minimum values are plotted as functions of the total number of non-zero elements estimated by the coordinate descent algorithm.
As can be seen, the coordinate descent algorithm performs best as it always returns the lowest objective function among the three methods.  The ECM converges to points that have slightly higher objective function  than \citet{BienTib2011} in dense scenario (Panel (c) and (d)), however, the difference is small (less than 0.05).  \citet{BienTib2011} seems to find points that are far less optimal than the coordinate descent or ECM algorithms in certain cases as is suggested by the spikes and dips in panel (a) and (b) where the differences in the objective function are larger than 1.

Finally, it is known that, for nonconvex problems, any optimization algorithms are not guaranteed to converge to a global minimum. It is often recommended to run algorithms at multiple initial values.  We then wish to compare the performance of the algorithms under different initial values.  In the previous experiments, all computations were initialized at the full sample covariance matrix $\bfSigma^{(0)} = \bS$. To be different, it is natural to initialize them at the other extreme case in which $\bfSigma^{(0)} = \textrm{diag}(s_{11},\ldots,s_{pp})$.  However, this initial value can only be used for the coordinate descent but not for ECM as discussed before, since the ECM iteration will stuck at $\bfSigma^{(0)}$ and never move when the off-diagonal elements are exactly equal to zero.  To avoid this problem, we start the ECM at $\textrm{diag}(s_{11},\ldots,s_{pp})+10^{-3}$ to avoid exact zeros on the off-diagonal elements.   For each of the four datasets, we picked three different values of $\rho$ so that they represent three different levels of sparsity based on the experiment before. For each of the four datasets and each of the three $\rho$s, we ran the three algorithms starting from the new set of initial values: $$\bfSigma^{(0)}_{CD} =\bfSigma_{BT}^{(0)} = \textrm{diag}(s_{11},\ldots,s_{pp}),\quad \bfSigma^{(0)}_{ECM} = \textrm{diag}(s_{11},\ldots,s_{pp})+10^{-3}.$$  We recorded the CPU time, sparsity of the minimum points and  the minimum value of the objective function. Table \ref{tab:1}
reports these measures.  Three things are worth noting. First,  \citet{BienTib2011}'s algorithm seems to get stuck at the initial value of a diagonal matrix in all cases. In contrast, the proposed algorithms work fine and find reasonable minimum points of $\bfSigma$ because the minimum values of the objective function and level of sparsity are quite close to those obtained from starting at the full sample covariance matrix.  We have tried to initialize \citet{BienTib2011}'s algorithm at $\bfSigma^{(0)}_{BT} = \textrm{diag}(s_{11},\ldots,s_{pp})+10^{-3}$ but found that it still gets stuck after a few iterations.  It may be possible to improve the poor performance of \citet{BienTib2011}'s algorithm  by adjusting some of its tuning parameters.  However, it may be safe to conclude that \citet{BienTib2011}'s algorithm requires either very careful tuning or performs badly at this important initial value.   Second, initial values indeed matter. Comparing the results between full and diagonal initial values, we see substantial differences in all three measures. In particular, the limiting points from the diagonal initial matrices are sparser than those from the full initial matrices. This is not surprising because of the drastic difference in sparsity between these two starting points.  Third, comparing the minimum values of the objective function achieved by the three algorithms (last two columns), we see that the coordinate descent often reaches the lowest minimum values regardless of the initial points. The few exceptions (e.g., sparse $\bfSigma$, $p=100,\rho=0.01$) when the coordinate descent is not the best seems to be the case that $\rho$ is small and the fraction of the number of non-zero elements is large.

\begin{table}[tbp]
\scriptsize
\center
\begin{tabular}{llrrrrrrrr}
\hline
 &  & \multicolumn{2}{c}{CPU Time} &  & \multicolumn{2}{c}{$\%$ Nonzero} &  & \multicolumn{2}{c}{Objective Func}\\
 Model&  Method& Full & Diag &  & Full & Diag &  & Full & Diag\\
\hline
\multirow{3}{*}{Sparse, $p=$100, $\rho=$0.01} & BT & 113 & 28 &  & 0.919 & 0.000 &  & 2.042 & 79.022\\

 & ECM & 14 & 21 &  & 0.918 & 0.912 &  & 2.030 & 2.030\\

 & CD & 135 & 102 &  & 0.920 & 0.918 &  & 2.033 & 2.032\\

&&&&&&&&& \\
\multirow{3}{*}{Sparse, $p=$100, $\rho=$0.24} & BT & 67 & 26 &  & 0.303 & 0.000 &  & 39.455 & 79.022\\

 & ECM & 26 & 122 &  & 0.316 & 0.256 &  & 39.463 & 39.504\\

 & CD & 23 & 19 &  & 0.304 & 0.302 &  & 39.454 & 39.454\\

&&&&&&&&& \\
\multirow{3}{*}{Sparse, $p=$100, $\rho=$1.11} & BT & 354 & 25 &  & 0.032 & 0.000 &  & 85.309 & 79.022\\

 & ECM & 114 & 5 &  & 0.032 & 0.000 &  & 85.310 & 79.022\\

 & CD & 21 & 0 &  & 0.032 & 0.000 &  & 85.306 & 79.022\\
&&&&&&&&& \\
&&&&&&&&& \\
&&&&&&&&& \\
\multirow{3}{*}{Sparse, $p=$200, $\rho=$0.01} & BT & - & 221 &  &  -& 0.000 &  &  -& 156.840\\

 & ECM & 300 & 368 &  & 0.880 & 0.877 &  & -4.407 & -4.407\\

 & CD & 2230 & 1452 &  & 0.883 & 0.886 &  & -4.403 & -4.404\\

&&&&&&&&& \\
\multirow{3}{*}{Sparse, $p=$200, $\rho=$0.15} & BT &  -& 173 &  &  -& 0.000 &  &  -& 156.840\\

 & ECM & 507 & 1486 &  & 0.311 & 0.281 &  & 54.941 & 54.971\\

 & CD & 536 & 371 &  & 0.309 & 0.307 &  & 54.931 & 54.931\\

&&&&&&&&& \\
\multirow{3}{*}{Sparse, $p=$200, $\rho=$1.00} & BT &  -& 170 &  &  -& 0.000 &  &  -& 156.840\\

 & ECM & 427 & 60 &  & 0.036 & 0.000 &  & 160.100 & 156.840\\

 & CD & 58 & 4 &  & 0.035 & 0.000 &  & 160.090 & 156.840\\

&&&&&&&&& \\
&&&&&&&&& \\
&&&&&&&&& \\
\multirow{3}{*}{Dense, $p=$100, $\rho=$0.02} & BT & 27 & 19 &  & 0.909 & 0.000 &  & 96.782 & 167.165\\

 & ECM & 14 & 17 &  & 0.912 & 0.890 &  & 96.784 & 96.785\\

 & CD & 45 & 18 &  & 0.909 & 0.880 &  & 96.783 & 96.782\\

&&&&&&&&& \\
\multirow{3}{*}{Dense, $p=$100, $\rho=$0.19} & BT & 55 & 25 &  & 0.313 & 0.000 &  & 152.914 & 167.165\\

 & ECM & 35 & 55 &  & 0.355 & 0.328 &  & 152.924 & 152.930\\

 & CD & 11 & 11 &  & 0.311 & 0.303 &  & 152.911 & 152.911\\

&&&&&&&&& \\
\multirow{3}{*}{Dense, $p=$100, $\rho=$0.32} & BT & 149 & 10 &  & 0.027 & 0.000 &  & 166.929 & 167.165\\

 & ECM & 73 & 19 &  & 0.057 & 0.000 &  & 166.956 & 167.169\\

 & CD & 23 & 17 &  & 0.024 & 0.009 &  & 166.910 & 166.953\\

&&&&&&&&& \\
&&&&&&&&& \\
&&&&&&&&& \\
\multirow{3}{*}{Dense, $p=$200, $\rho=$0.01} & BT & 330 & 147 &  & 0.928 & 0.000 &  & 180.274 & 342.279\\

 & ECM & 260 & 325 &  & 0.924 & 0.914 &  & 180.277 & 180.280\\

 & CD & 342 & 222 &  & 0.928 & 0.927 &  & 180.275 & 180.274\\

&&&&&&&&& \\
\multirow{3}{*}{Dense, $p=$200, $\rho=$0.15} & BT & 657 & 207 &  & 0.294 & 0.000 &  & 303.813 & 342.279\\

 & ECM & 610 & 764 &  & 0.321 & 0.317 &  & 303.821 & 303.823\\

 & CD & 95 & 95 &  & 0.292 & 0.294 &  & 303.803 & 303.803\\

&&&&&&&&& \\
\multirow{3}{*}{Dense, $p=$200, $\rho=$0.29} & BT & 890 & 139 &  & 0.034 & 0.000 &  & 341.231 & 342.279\\

 & ECM & 1154 & 263 &  & 0.060 & 0.000 &  & 341.255 & 342.285\\

 & CD & 171 & 225 &  & 0.032 & 0.026 &  & 341.221 & 341.223\\
\hline
\end{tabular}
\center
\caption{Performance of three algorithms starting at two different initial values: Full, $\bfSigma^{(0)}=\bS$; and Diag, $\bfSigma^{(0)}=\textrm{diag}(s_{11},\ldots,s_{pp})$.  The ``CPU Time'' columns
present the CPU run time in seconds;  the ``\% Nonzero'' columns
present the percentage of nonzero elements in the minimum points;  the ``Objective Func'' columns
present the minimum value of the objective function. The three algorithms are: \citet{BienTib2011} (BT), the Expectation/Conditional Maximization algorithm (ECM) of Section \ref{sec:ecm}  and the coordinate descent (CD) of Section \ref{sec:cd}.  }
\label{tab:1}
\end{table}

\section{Discussion}
We have developed two alternative algorithms for fitting sparse covariance graphical lasso models using $L_1$ penalty. These two algorithms are shown to be much easier to implement, significantly faster to run and numerically more stable than the algorithm of \citet{BienTib2011}.   Both MATLAB and R software packages implementing the new algorithms for solving covariance graphical models are freely available from the author's the website of the paper.

\bibliography{CovGlasso}

\begin{thebibliography}{10}
\expandafter\ifx\csname natexlab\endcsname\relax\def\natexlab#1{#1}\fi

\bibitem[{{Armagan} et~al.(2011){Armagan}, {Dunson} \& {Lee}}]{Armagan11}
\textsc{{Armagan}, A.}, \textsc{{Dunson}, D.} \& \textsc{{Lee}, J.} (2011).
\newblock {Generalized double Pareto shrinkage}.
\newblock \textit{Statistica Sinica (forthcoming)} .

\bibitem[{Bien \& Tibshirani(2011)}]{BienTib2011}
\textsc{Bien, J.} \& \textsc{Tibshirani, R.~J.} (2011).
\newblock Sparse estimation of a covariance matrix.
\newblock \textit{Biometrika} \textbf{98}, 807--820.

\bibitem[{{Breheny} \& {Huang}(2011)}]{BrehenyHuang2011}
\textsc{{Breheny}, P.} \& \textsc{{Huang}, J.} (2011).
\newblock {Coordinate descent algorithms for nonconvex penalized regression,
  with applications to biological feature selection}.
\newblock \textit{The Annals of Applied Statistics} , 232--253.

\bibitem[{Dempster(1972)}]{Dempster72}
\textsc{Dempster, A.} (1972).
\newblock Covariance selection.
\newblock \textit{Biometrics} \textbf{28}, 157--75.

\bibitem[{Friedman et~al.(2007)Friedman, Hastie, Höfling \&
  Tibshirani}]{FHHT2007}
\textsc{Friedman, J.}, \textsc{Hastie, T.}, \textsc{Höfling, H.} \&
  \textsc{Tibshirani, R.} (2007).
\newblock Pathwise coordinate optimization.
\newblock \textit{The Annals of Applied Statistics} \textbf{1}, pp. 302--332.

\bibitem[{Friedman et~al.(2008)Friedman, Hastie \& Tibshirani}]{Friedman08}
\textsc{Friedman, J.}, \textsc{Hastie, T.} \& \textsc{Tibshirani, R.} (2008).
\newblock {Sparse inverse covariance estimation with the graphical lasso}.
\newblock \textit{Biostatistics} \textbf{9}, 432--441.

\bibitem[{Meng \& Rubin(1993)}]{MengRubin1993}
\textsc{Meng, X.-L.} \& \textsc{Rubin, D.~B.} (1993).
\newblock Maximum likelihood estimation via the ecm algorithm: A general
  framework.
\newblock \textit{Biometrika} \textbf{80}, 267--278.

\bibitem[{{Polson} \& {Scott}(2011)}]{PolsonScott2011}
\textsc{{Polson}, N.~G.} \& \textsc{{Scott}, J.~G.} (2011).
\newblock {Data augmentation for non-Gaussian regression models using
  variance-mean mixtures}.
\newblock \textit{ArXiv e-prints} .

\bibitem[{Tseng(2001)}]{Tseng2001}
\textsc{Tseng, P.} (2001).
\newblock Convergence of a block coordinate descent method for
  nondifferentiable minimization.
\newblock \textit{Journal of Optimization Theory and Applications}
  \textbf{109}, 475--494.

\bibitem[{Wu \& Lange(2008)}]{WuLange2008}
\textsc{Wu, T.~T.} \& \textsc{Lange, K.} (2008).
\newblock Coordinate descent algorithms for lasso penalized regression.
\newblock \textit{The Annals of Applied Statistics} \textbf{2}, pp. 224--244.

\end{thebibliography}
\bibliographystyle{biometrika}
\end{document}